# Titanic overconfidence—dark uncertainty can sink hybrid metrology for semiconductor manufacturing

Ronald G. Dixson, Adam L. Pintar, R. Joseph. Kline, Thomas A. Germer, J. Alexander Liddle, John S. Villarrubia, and Samuel M. Stavis

Hybrid metrology for semiconductor manufacturing is on a collision course with dark uncertainty. An IEEE technology roadmap for this venture has targeted a linewidth uncertainty of ± 0.17 nm at 95 % coverage and advised the hybridization of results from different measurement methods to hit this target. Related studies have applied statistical models that require consistent results to compel a lower uncertainty, whereas inconsistent results are prevalent. We illuminate this lurking issue, studying how standard methods of uncertainty evaluation fail to account for the causes and effects of dark uncertainty. We revisit a comparison of imaging and scattering methods to measure linewidths of approximately 13 nm, applying contrasting statistical models to highlight the potential effect of dark uncertainty on hybrid metrology. A random effects model allows the combination of inconsistent results, accounting for dark uncertainty and estimating a total uncertainty of ± 0.8 nm at 95 % coverage. In contrast, a common mean model requires consistent results for combination, ignoring dark uncertainty and underestimating the total uncertainty by as much as a factor of five. To avoid such titanic overconfidence, which can sink a venture, we outline good practices to reduce dark uncertainty and guide the combination of indeterminately consistent results.

*Index Terms*—consensus, hierarchical, inter-comparison.

## I. INTRODUCTION

THE fundamental purpose of estimating uncertainty is to quantify confidence and guide decisions. An accurate estimate of uncertainty helps to quantify the risk and reward of possible outcomes and make rational plans. An underestimate of uncertainty causes overconfidence, which can misguide decisions, leading to catastrophe and sinking a venture. Narrowing the scope of the problem to measurement results, which guide decisions in many scientific and technological ventures,[1] unknown errors cause underestimates of uncertainty. This problem can come to light, however, in comparisons of results from different tools, methods, and laboratories. Different methods are sensitive to different properties of physical samples, so inconsistency among the results of inter-method comparisons may be unsurprising. But even when the participants of inter-laboratory comparisons apply the same method, the variability among the results commonly exceeds the uncertainty estimate for each result. The same can apply to inter-tool variability within one method.

Inconsistent results are a persistent problem in measurements of physical constants[2] and material properties[3], as well as in clinical trials[4]. A meta-analysis of results from the field of analytical chemistry[5] described the excess variability as dark uncertainty. This apt term is reminiscent of dark matter, implying its invisibility and abundance. Even ordinary errors can be invisible. For example, measurements are imperfectly repeatable, but without replicate measurements to manifest the variability of the results, random effects that limit measurement repeatability are invisible. Systematic effects from unknown causes pose a less obvious and more serious problem. A bias particular to a single method, for example, remains dark within the results of that method absent a correction.

Ideally, the standard process of evaluating uncertainty would identify all of its causes, including measurement repeatability and correction uncertainty. The estimate of all of the effects on the result would approximate the total uncertainty. In reality, the total uncertainty exceeds the uncertainty estimate by an amount that corresponds to the dark uncertainty. Multiple causes in various categories can contribute to dark uncertainty (Figure 1), which is quantifiable in a comparison but unattributable to any particular cause.

In this study, we warn that semiconductor manufacturing is on a collision course with dark uncertainty. Linewidth or critical-dimension (CD) measurements pervade this venture, with the results informing decisions throughout process development, factory startup, yield engineering, and cost reduction[6]. To guide these tasks, an IEEE technology roadmap has targeted an uncertainty of ± 0.17 nm at 95 % coverage for the widths of isolated lines by 2028[6]. Noting that this small target is hard to hit by any individual measurement process, this influential guide has advised the use of hybrid metrology.

Ronald G. Dixson is with NIST, Gaithersburg, MD 20899 USA (email: ronald.dixson@nist.gov).
Adam L. Pintar is with NIST, Gaithersburg, MD 20899 USA (email: adam.pintar@nist.gov).
R. Joseph Kline is with NIST, Gaithersburg, MD 20899 USA (email: r.kline@nist.gov).
Thomas A. Germer is with NIST, Gaithersburg, MD 20899 USA (email: thomas.germer@nist.gov).
J. Alexander Liddle is with NIST, Gaithersburg, MD 20899 USA. (email: james.liddle@nist.gov).
John S. Villarrubia is with NIST, Gaithersburg, MD 20899 USA (email: john.villarrubia@nist.gov).
Samuel M. Stavis is with NIST, Gaithersburg, MD 20899 USA (email: samuel.stavis@nist.gov).
Corresponding author: Samuel M. Stavis.

The authors performed this work with funding support from the CHIPS Metrology Program, part of CHIPS for America, NIST, United States Department of Commerce.

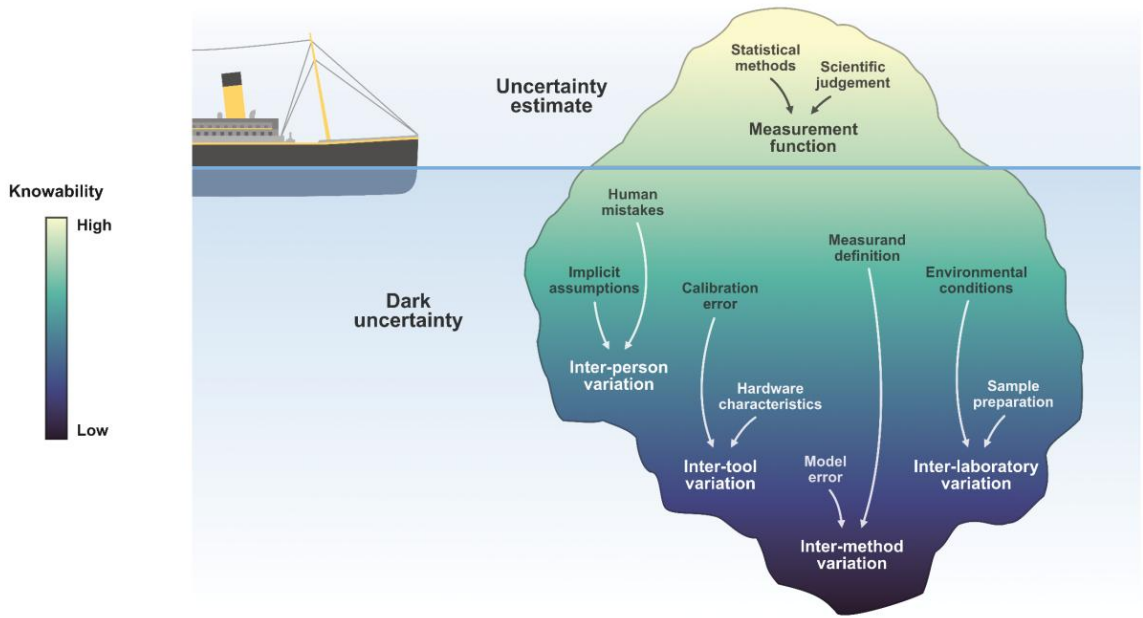

**Fig. 1.** Iceberg chart comparing (above water) an uncertainty estimate to (under water) dark uncertainty. The standard process of evaluating uncertainty yields an underestimate of total uncertainty. Dark uncertainty correlates with categories of observation.

The meaning of this term is itself vague, having been used loosely to describe various analyses. We weight the primary literature[7-10] heavily to define hybrid metrology as the statistical combination of results from different measurement processes. The best argument for hybrid metrology is that it can leverage the complementary strengths of dissimilar principles of measurement to reduce their combined uncertainty. Early efforts to achieve this benefit applied Bayesian priors[7] or combined regression[10]. The prior implementations of these statistical models required consistency of the combined results. The primary literature acknowledged this requirement[7-10] but later reviews and summaries[11-13] disregarded its significance, as have emerging applications[14-16], all while touting the promise of hybrid metrology to achieve a lower uncertainty. None of these studies has confronted the prevalence of inconsistent results.

As linewidths have decreased through the nanometer scale, measurements have also advanced. Metrologists have long known of the potential for inconsistent results from different methods[17-20], motivating the application and comparison of multiple methods[21,22]. Individually or in combination, optical microscopy (OM)[23,24], scanning electron microscopy (SEM)[25-27], electrical resistance linewidth metrology[28-30], atomic force microscopy (AFM)[31], scatterometry[32,33], transmission electron microscopy (TEM)[34-37], and X-ray scattering[38] have all been relevant to semiconductor manufacturing. At present, the most prevalent methods for high-volume manufacturing (HVM) are critical-dimension SEM (CD-SEM) and scatterometry, or optical critical-dimension (OCD). Critical-dimension small-angle X-ray scattering (CD-SAXS) is increasingly important, and critical-dimension AFM (CD-AFM) remains relevant[39].

We study causes and effects of dark uncertainty among these methods, revisiting several inter-comparisons of linewidth measurements. We focus on TEM, SEM, and CD-SAXS results[20] of approximately 13 nm, which range from marginally consistent to slightly inconsistent. Two contrasting statistical models of decreasing believability[40] highlight the quantification of dark uncertainty and its potential effect on hybrid metrology. More believably, a random effects model allows the combination of inconsistent results[41], recognizing both the uncertainty estimate for each result and the excess variability that becomes evident from the comparison of results. This statistical model estimates a consensus mean with a total uncertainty of ± 0.8 nm at 95 % coverage, as well as a dark uncertainty of 0.5 nm. Less believably, a common mean model[42] requires consistent results, combining independent realizations of the same measurand without bias. This statistical model ignores dark uncertainty and compels an uncertainty as much as five times lower than the total uncertainty.

These results illuminate dark uncertainty as a lurking issue for hybrid metrology, which can yield such a deep underestimate of total uncertainty as to cause titanic overconfidence. In light of this issue, we outline good practices to evaluate uncertainty, provide a study chart to combine results of indeterminate consistency, and comment on future studies to improve consistency. Our point of view facilitates better—but often higher—estimates of uncertainty from combinations of results in semiconductor manufacturing and other ventures.

## II. DARK UNCERTAINTY HIDES FROM EVALUATION

We explain how dark uncertainty hides from evaluation. Our explanation is sufficiently formal yet pedagogical to begin to understand this subtle problem, therefore superseding the explanation of uncertainty in the IEEE technology roadmap[6].

### A. Measurand Definition

The measurand is the quantity intended to be measured. A direct comparison of results requires the same measurand. However, each measurement process reports the value of its realized quantity[43,44]. Different measurement processes can realize different quantities that go by the same name, leading to questions about measurand definition. For example, is the linewidth the value at the top, bottom, or some kind of average? In measurements of a grating, what is the averaging effect of scattering methods such as OCD and CD-SAXS relative to imaging methods such as AFM and SEM? In inter-National Metrology Institute (NMI) comparisons, participants attempt to agree on a measurand definition and account for all corrections and uncertainties in linking their practical measurements to the abstract definition, making a conscious effort to ensure that the realized quantity is the same as the measurand. An inadequately specific definition gives an ambiguous measurand that admits a range of correct values. A previous study addressed this issue in dimensional metrology at larger scales[19].

### B. Uncertainty Evaluation

An NMI advises evaluating uncertainty by a cause-and-effect analysis[5,44,45], with several underappreciated issues. A measuring tool produces a signal $S$, which is a function $S = P(m; q_i)$ of the measurand $m$, as well as influence quantities $q_i$ with index $i = 1, 2, \ldots, N$ that also affect the signal. The signal could be as simple as a scalar value but is more commonly multidimensional, such as an image or diffraction pattern. The signal and influence quantities are inputs to the measurement function $f: m = f(S; q_i)$ with $f$ the inverse function of $P$. The Guide to the Expression of Uncertainty in Measurement (GUM)[44] notes that $f$ can be an analytic expression or can involve numerical evaluation. For example, the value of $m$ and values of $q_i$ may be possible to determine by fitting the function $P$ to the signal $S$.

In an additive error model, influence quantities vary randomly around a non-zero mean $q_i = \hat{q}_i + \lambda_i + \varepsilon_{ij}$ with $\hat{q}_i + \lambda_i$ the mean of the $i^{\text{th}}$ influence quantity—$\hat{q}_i$ is its estimate and $\lambda_i$ is a constant correction for estimate error—and $\varepsilon_{ij}$ corrects a random error of zero mean at the $j^{\text{th}}$ measurement. To first order

$$m = f(S, \hat{q}_i) + (\varepsilon_T + \lambda_T)\frac{\partial f}{\partial q_T}\bigg|_{q_T = \hat{q}_T} + (\varepsilon_M + \lambda_M)\frac{\partial f}{\partial q_M}\bigg|_{q_M = \hat{q}_M} + (\varepsilon_L + \lambda_L)\frac{\partial f}{\partial q_L}\bigg|_{q_L = \hat{q}_L} + \ldots \quad (1)$$

In Equation 1, the terms $\varepsilon$ and $\lambda$ are corrections for the respective effects of random and systematic errors in $\hat{q}_i$ that are attributable to tools $T$, methods $M$, and laboratories $L$. The ellipsis denotes similar terms due to errors of $S$ and other influence quantities. Estimation of the value of $m$ by replacement of the random errors with the corresponding mean values, which are zero by construction, gives $\hat{m} = f(S, \hat{q}_i)$.

Treatment of errors as independent and random leads, through Equation 1 and the definition of variance, to

$$u_m^2 = R^2 + \left(\frac{\partial f}{\partial q_T}\right)^2 u_T^2 + \left(\frac{\partial f}{\partial q_M}\right)^2 u_M^2 + \left(\frac{\partial f}{\partial q_L}\right)^2 u_L^2 + \cdots \quad (2)$$

In Equation 2, $u_m$ is the standard uncertainty of the measurand value, $R^2$ is the repeatability variance of a time series of replicate measurements of $m$, $u_T^2$ is the variance of the $\lambda_T$ distribution, and likewise for $u_M^2$ and $u_L^2$. The repeatability variance

$$R^2 = \left(\frac{\partial f}{\partial q_T}\right)^2 \langle \varepsilon_T^2 \rangle + \left(\frac{\partial f}{\partial q_M}\right)^2 \langle \varepsilon_M^2 \rangle + \left(\frac{\partial f}{\partial q_L}\right)^2 \langle \varepsilon_L^2 \rangle + \cdots \quad (3)$$

combines all random errors, including the $\varepsilon$ terms.

In Equation 3, $R^2$ is measurable even if $f$ or the identities of the influence quantities are unknown. Therefore, $R^2$ is an unlikely hiding place for dark uncertainty at the time scale of the replicate series. The other terms are estimates from state-of-knowledge distributions, which are more challenging to quantify and can change only by more knowledge. Our notation in Equations 1–3 refrains from identifying terms for specific influence quantities in favor of categories with which the quantities correlate. We do so advisedly, as an influence often comes to light by observation of a correlation that informs us of the existence of the influence but not of its identity. An expression such as Equation 2 is then the best that is possible. Incognito influence quantities are good hiding places for excess uncertainty, suggesting an aphorism for dark uncertainty—*Incognito, ergo sum.*

### C. Component Estimation

As a conceptual representation of the corresponding effect of dark uncertainty, we introduce the Iceberg Chart of Uncertainty Evaluation (IceCUE) (Figure 2a). IceCUE shows uncertainty components as variance icebergs, with height above water indicating a component estimate and depth below water indicating the relative amount of dark uncertainty. We depict equal variance components for simplicity. Omission of the underwater parts from the sum of variances causes an underestimate of total uncertainty.

### III. CAUSES AND EFFECTS OF DARK UNCERTAINTY

We partition causes of dark uncertainty into four categories (Figure 1) and compare approaches to treating its effects.

### A. Inter-person Variation

Metrologists make judgments, assumptions, and even mistakes. They strive to model measurement processes and interpret measurement results. Uncertainty evaluation by other-than-statistical methods relies heavily on scientific judgment with supporting information, and both should be explicit and available for review. Implicit assumptions could be difficult to detect. Overconfidence bias varies from person to person, depending in part on scientific knowledge[46].

### B. Inter-tool Variation

Incomplete understanding of how influence quantities affect measuring tools is a problem in ongoing efforts to match results from different tools in semiconductor manufacturing.

For example, failure to recognize the effect of electron beam tilt, $\theta_t$, in an SEM measurement of linewidth means that $f$ in the measurement model lacks a correction for $\theta_t$, and Equation 2 lacks a $u_{\theta_t}$ term. In this way, the tilt effect escapes correction or coverage by an uncertainty estimate. If SEM tools A and B have different tilts, the same feature produces different signals with measurand values $f(S_A)$ and $f(S_B)$, revealing the existence but not the identity of an influence quantity. Any influence common to the tools and results is unknown.

Electron beam shape in SEM is another influence quantity[47], as are aberrations[48] and misalignments[49] in optical microscopy. To compensate, manufacturers focus on repeatability metrics and choose offsets to reconcile inconsistent results. However, adding a constant value to a consistent set of offsets produces a new consistent set. Without knowing which set is correct, dark uncertainty remains. As well, the stability of an offset depends on that of unknown influence quantities. Drift of an offset can become increasingly problematic for tighter tolerances.

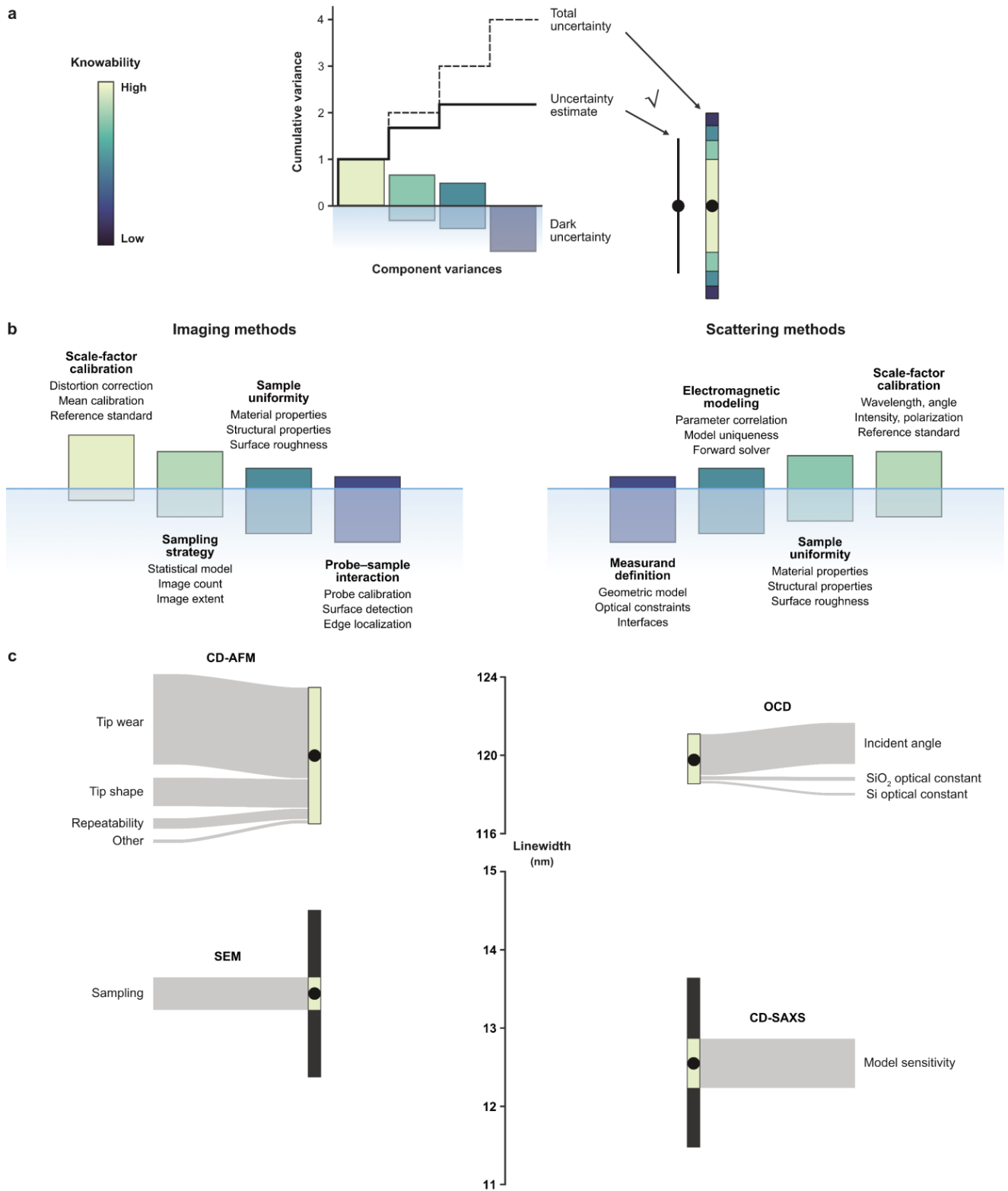

**Fig. 2.** (a) Iceberg Chart of Uncertainty Evaluation (IceCUE) showing uncertainty components as variance icebergs with deepening underestimates. (b) Iceberg charts showing four categories of uncertainty components for imaging methods such as CD-AFM or SEM, and for scattering methods such as OCD or CD-SAXS. (c) Sankey charts showing dominant uncertainty components for CD-AFM–OCD[49] and SEM–CD-SAXS[20] inter-comparisons. Bar heights are proportional to fractional variances. The dark uncertainty estimate is per Reference 4. A more complete analysis is in Table 1, Table 2, and Figure 3.

*C. Inter-laboratory Variation*

Semiconductor manufacturers seek to match results from different sites. Similarly, the Consultative Committees (CCs) of the International Committee on Weights and Measures (CIPM) perform inter-NMI comparisons to establish metrological equivalence. To the extent that tools and methods differ between laboratories, the laboratories subsume the resulting variability. Influence quantities such as environmental factors and local standards are also laboratory-dependent, affecting inter-laboratory variation in the same way that $\theta_t$ affects inter-tool variation.

Many inter-NMI comparisons have applied an analysis[50] that excludes inconsistent results to reduce the uncertainty of a weighted average, which serves as a reference value. We consider a random effects model, which recognizes excess variability and estimates dark uncertainty, to be more appropriate.

*D. Inter-method Variation*

A study of inter-method variation begins with the measurand. A simplistic definition of a linewidth measurand—the distance between the left and right edges of a line—is unambiguous for a uniform feature with a rectangular cross-section. However, real features deviate from this idealization. For a trapezoidal cross-section, the edge separation varies from top to bottom, requiring either an average or choice of particular height in the measurand definition, affecting different methods differently.

Iceberg charts identify potential causes of dark uncertainty among imaging and scattering methods (Figure 2b). Some uncertainties, such as those of reference standards for scale-factor calibration[48], are quite knowable. Even so, differences between a standard and a sample could result in different interactions with a measuring tool, causing bias. Uncertainties attributable to measurand definitions and probe–sample interactions are less knowable. Measurand definition is a subtle problem for scattering methods, which involve measurements across larger areas than imaging methods and make corresponding assumptions, such as the absence of systematic variation of feature shape across the area.

Each method requires a model of the probe–sample interaction. For example, CD-SEM requires a model of the relationship between the electron signal and surface topography. OCD or CD-SAXS signals depend on optical or X-ray scattering, and the corresponding models involve different physics. Finite knowledge of the interaction physics leads to dark uncertainty in the functional form of $f$ in Equations 1 and 2, requiring judgment of which forms of $f$ are physically reasonable. As well, models often include geometric parameters that can fail to describe the true shapes of features.

Manufacturing processes yield variable linewidth, which is the measurand, and shape parameters, which are influence quantities[51,52]. Both the measurand and the influence quantities affect the signal. If the effects are possible to distinguish, then the contributions are possible to separate in a multiparameter fit of $S$ by $P(m; q_i)$ with $m$ and the $q_i$, including shape parameters, as fitting parameters. In contrast, different groups of parameters that affect the signal similarly have a correlation, and their contributions are difficult to disambiguate.

This problem of parametric correlation motivates hybrid metrology[7], which is an optimistic—possibly over-optimistic—type of inter-method comparison. An early example compared AFM and OCD results for a trapezoidal cross-section[7]. The AFM results are more certain than the OCD results, motivating the use of the former to constrain the latter. As the authors of this study realized[7], the hybridization requires consistent results to compel a lower uncertainty than either individual result. Yet, the bottom of three trapezoidal linewidths were inconsistent, while the heights were inconsistent prior to assumption of an oxide layer. Similar issues have arisen in other studies[8,9,14-16].

As each method uses its own model, corresponding errors and uncertainties are dark within each method. NIST has studied this issue by comparing methods of electrical resistance linewidth, AFM, and SEM[22], top-down SEM with SEM in cross-section[53], CD-AFM and OCD[49], and SEM, TEM, and CD-SAXS[20]. Paradoxically, as these methods have improved, the reduction of uncertainty estimates has increased the relative impact of dark uncertainty, such as from model errors. For AFM, SEM, and electrical resistance linewidth[22], and for CD-AFM and OCD middle linewidth[49], the uncertainty estimates are large enough to obscure any dark uncertainty (Figure 2c), but for SEM and CD-SAXS the results are inconsistent (Figure 2c). While many causes of uncertainty are possible, Sankey charts for two inter-method comparisons—CD-AFM–OCD[49] and SEM–CD-SAXS[20]—show only a few major components of uncertainty (Figure 2c). The uncertainty evaluations depend on the inter-comparison purpose. The CD-AFM–OCD comparison aimed for independently traceable results, achieving apparent consistency for relatively large linewidths[49]. The SEM–CD-SAXS comparison aimed to isolate potential causes of dark uncertainty by reducing the number of uncertainty components for each method[20], motivating further study. In the SEM–CD-SAXS comparison, different people evaluated the uncertainty of measurement results from different tools, methods, and laboratories, so that all of the categories are potential correlates of dark uncertainty. In Equation 1, an erroneous $\hat{q}$ leads to a non-zero $\lambda$ and systematic error of $\lambda \frac{\partial f}{\partial q}\big|_{q=\hat{q}}$, contributing to the SEM–CD-SAXS inconsistency.

TABLE 1
MIDDLE LINEWIDTHS AND UNCERTAINTIES FROM REFERENCE 20. UNCERTAINTIES ARE 95% COVERAGE INTERVALS.

|  | TEM | SEM | CD-SAXS |
|---|---|---|---|
| Experimental measurement result | 12.9 nm ± 0.3 nm | 13.4 nm ± 0.2 nm | 12.6 nm ± 0.3 nm |

TABLE 2
RANDOM EFFECTS MODEL OF MIDDLE LINEWIDTHS AND UNCERTAINTIES FROM REFERENCE 20.

|  | Consensus linewidth | 95% coverage interval for consensus linewidth | Dark uncertainty | 95% coverage interval for dark uncertainty |
|---|---|---|---|---|
| Random effects model | 13.0 nm | 12.2 nm to 13.8 nm | 0.5 nm | 0.2 nm to 1.3 nm |

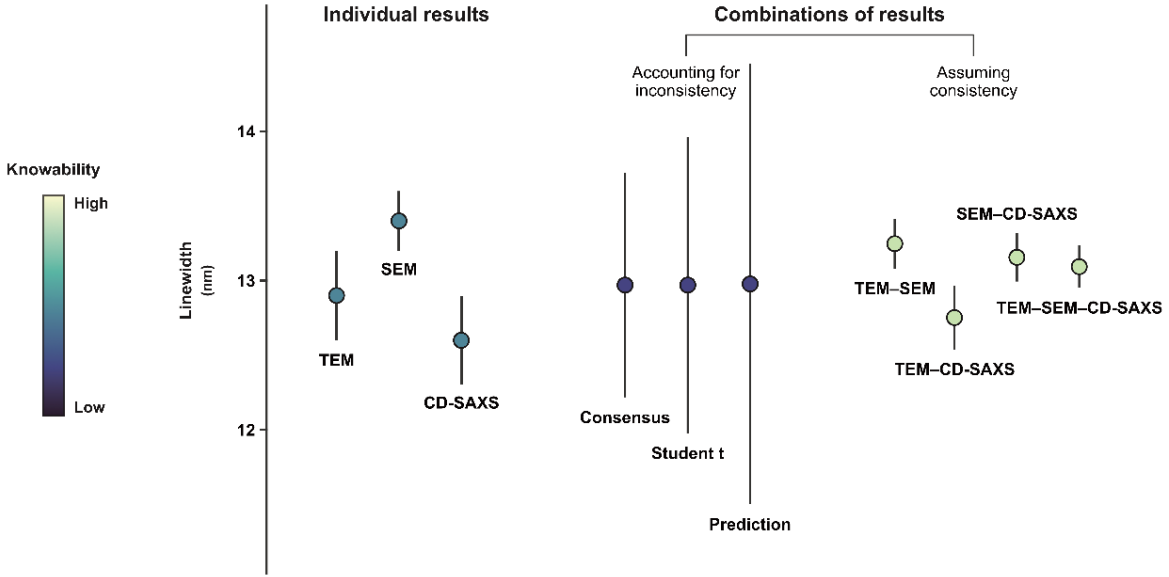

**Fig. 3.** Plot comparing mean values and 95 % coverage intervals in three categories. The first category includes individual results from different measurement processes[20]. The second category includes combinations of results that account for inconsistency, most properly by a random effects model that determines a consensus value and accounts for dark uncertainty. The third category includes combinations of results that assume consistency to compel a lower uncertainty by a common mean model.

If errors systematic to a method are random and independent among different methods, then the mean value of the results from different methods is a better estimate of the true value than the individual results, even ignoring any potential errors in common. The same idea applies to tools and laboratories. The random effects model elaborates on this idea.

## IV. EVALUATION OF DARK UNCERTAINTY

To illustrate how dark uncertainty can sink hybrid metrology, we revisit a previous comparison[20] of TEM, SEM, and CD-SAXS measurements of linewidths of approximately 13 nm (Table 1, Figure 3). Features of this size are still relevant to current technology as the test sample was technologically advanced at the time of the comparison. Similarly, the SEM and CD-SAXS methods of this study remain near or even surpass current methods for HVM, whereas TEM remains costly and slow but returns valuable data. To isolate dark uncertainty from potential causes such as model and definitional errors, this comparison removed confounding components of uncertainty, such as of scale factor, by normalization.

### A. Random Effects Model

We perform[54] a new analysis that allows for inconsistency of these results and recognizes any dark uncertainty (Table 2, Figure 3). A random effects model, like that of the NIST Consensus Builder[41], accounts for the uncertainty estimate of each result, as well as any excess variance among the results.

Without knowing the influence quantities, we cannot apply Equation 2 directly. Instead, we can estimate an uncertainty that is large enough to cover the results. Due to the possibility of errors that occur in common, the dark uncertainty is a lower bound of the combination of $u_T$, $u_M$, and $u_L$ in Equation 2. In either a Bayesian or frequentist context, an estimate of dark uncertainty can reach zero, implying consistent results. If the lower bound of a 95 % coverage interval for the estimate of dark uncertainty exceeds zero, as is the case in Table 2, the results are likely to be inconsistent. Two other tests yield similar conclusions—two of the three comparisons of method pairs have 95 % coverage intervals that fail to overlap, while the Cochran Q test[55] for inconsistency yields a p-value below 0.01.

### B. Common Mean Model

We perform another new analysis of these results that requires consistency and ignores dark uncertainty (Figure 3), making it a toy model of prior implementations of hybrid metrology. The common mean model combines each result as an independent realization of the same measurand without bias—a sample from a normal distribution with a common unknown mean, assuming a flat improper prior, and a different known variance[42]. Prior implementations of hybrid metrology intended to benefit from the complementarity of different measurement processes to achieve a lower uncertainty. Although simpler in intention, the common mean model is mathematically similar to using the results as Bayesian priors[7], and similarly compels a lower uncertainty of the combination of results relative to the uncertainty of each result[42].

### C. Uncertainty Comparison

The combination of a few inconsistent results should increase the total uncertainty due to dark uncertainty (Figure 3). Even if two methods yield marginally consistent results, such as TEM and CD-SAXS, the assumption of zero dark uncertainty is perilous. The hybrid TEM–CD-SAXS result is inconsistent with the SEM,

BOX 1

> **Good practices for uncertainty evaluation
> in semiconductor manufacturing and beyond**
>
> Although measurement challenges vary with specific applications, the general concepts that we outline in this point of view illustrate some good practices. To summarize, these are:
>
> 1. **Define ambiguous terms such as resolution,** considering that measurement resolution, the minimum detectable difference, is often more relevant than imaging resolution, the minimum resolvable separation.
>
> 2. **Define the measurand,** considering the demands of the application, the capabilities of the available methods and tools, and the potential inconsistency of the intended measurand and realized quantity.
>
> 3. **Define the measurement function,** preferably as an analytic expression or otherwise as an algorithm, as well as the uncertainty distribution and parameter values for each input term of the measurement function.
>
> 4. **Identify influence quantities and systematic effects** that cause measurement errors. Correct systematic effects in the measurement function, unless the correction uncertainty exceeds the error.
>
> 5. **Consider reasonable deviations** from the measurement function and uncertainties, such as an asymmetric distribution instead of a symmetric distribution for an uncertainty component.
>
> 6. **Use graphical schema** such as flow charts to understand measurement processes, iceberg charts to consider uncertainty components, and study charts to guide the combination of measurement results.
>
> 7. **Propagate uncertainty by Monte-Carlo methods** that specify all distributional assumptions and are exact up to Monte-Carlo error. Compare the results to the law of propagation of uncertainty.
>
> 8. **Consider both spatial and temporal effects** in a sampling strategy. Measurements that occur within a short time or at target positions within a short distance of each other may exhibit correlation.
>
> 9. **Evaluate the reasonableness of a statistical model** to describe experimental results for unceratinty evaluation. At least, compare simulation results from the statistical model to experimental results.
>
> 10. **Weigh the consistency of the results,** such as by estimating dark uncertainty using a random effects model, before combining results by a statistical model that requires consistency to decrease total uncertainty.

hybrid TEM–SEM, hybrid SEM–CD-SAXS, and hybrid TEM–SEM–CD-SAXS results, ignoring dark uncertainty that the SEM result brings to light. In contrast, the total uncertainty from the random effects model exceeds the individual result uncertainties, covers the excess variability among the results, and is up to five times larger than the uncertainty from the common mean model.

*D. Other Intervals*

A simple alternative to the random effects model is a Student $t$ model (Figure 3), which treats the results as replicate samples from a common normal distribution with an unknown mean and variance. This statistical model accounts for dark uncertainty by the variance estimate but weights the results equally, which is simplistic if the individual uncertainties differ. We also include a prediction interval from the random effects model for a fourth measurement process, such as CD-AFM, assuming that it manifests a deviation similar in magnitude to the first three. The prediction interval from the random effects model is up to ten times larger than the uncertainty from the common mean model, further emphasizing the risk of titanic overconfidence.

V. REDUCTION OF DARK UNCERTAINTY

*A. Good Practices*

Building on efforts to close gaps between abstract concepts and concrete applications in guides to uncertainty evaluation[45], we present ten good practices (Box 1), embodying the practice of using graphical schema[56,57] to improve understanding.

*B. Combining Results*

Expanding on this good practice, we introduce the Dark Uncertainty Study Chart (DUSC) to guide the combination of individual results (Figure 4). DUSC has a similar root as the NIST Decision Tree[58] but presents new branches to study dark uncertainty and advises the assumption of inconsistency for a few results without evidence to support their consistency.

DUSC begins with a test for dark uncertainty, the presence of which motivates review of each cause-and-effect evaluation of uncertainty. Correction of latent errors and propagation of higher uncertainty can decrease dark uncertainty. However, in the absence of any new insight, or if dark uncertainty remains even after rational revision, then a statistical model that allows the combination of inconsistent results provides a better estimate of total uncertainty than a statistical model that requires consistent results to compel a lower uncertainty. The NIST Decision Tree could recommend a particular random-effects model to complete the combination of results.

The ad hoc inflation of uncertainty to force consistency for hybridization is inadvisable, as inconsistency tests can fail before the inflated uncertainty reaches the dark uncertainty. For the results in Table 1, the Cochran Q test fails to detect inconsistency at a p-value above 0.1 upon inflation of each squared standard uncertainty by $(0.25 \text{ nm})^2$, which is a quarter

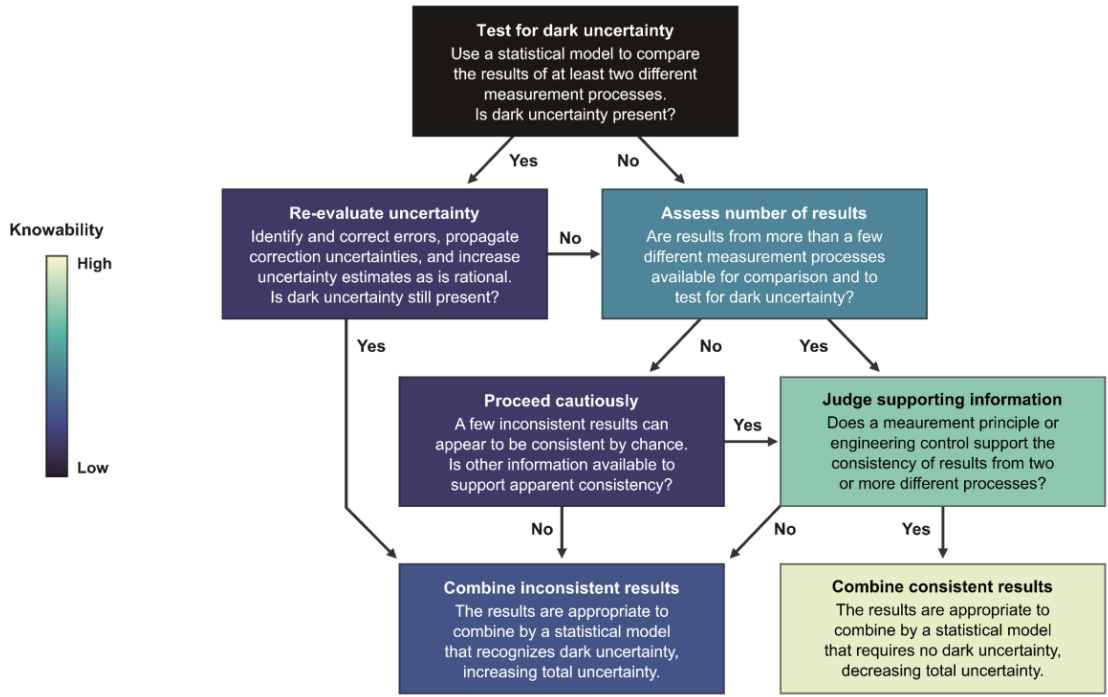

**Fig. 4.** Dark Uncertainty Study Chart (DUSC) guiding the combination of results of indeterminate consistency. DUSC tests for dark uncertainty, guides its potential reduction, and seeks information to support consistency before combining results.

of the necessary amount of $(0.5 \text{ nm})^2$ due to dark uncertainty.

Assessing the consistency of a few results is challenging. While extreme inconsistency is obvious, dark uncertainty that is comparable to the uncertainty estimate of an individual result precludes the use of a statistical model that requires consistency to compel a lower uncertainty. Even if a few results appear to be consistent, the notion that the corresponding measurement processes have realized the same quantity is questionable. The inherent limitation of comparing only a few results is low statistical power due to high sampling variability. Measurement principles[59] or engineering controls such as check standards and control charts could support an argument for consistency.

The availability of many results from different measurement processes is highly beneficial, allowing the total uncertainty to decrease below the uncertainty estimates of individual results in a random effects model[60]. Unlike statistical models that require consistency to compel a lower uncertainty, this decrease of uncertainty occurs in the presence of dark uncertainty, as effectively random deviations tend to cancel. This cancellation motivates the development of test structures for semiconductor manufacturing that are widely measurable for inter-comparison, which could support the certification of standards that support hybrid metrology by the study of influence quantities.

*C. Influence Quantities*

In Equations 1–3, we describe terms in categories of observations, because the influence quantities are unknown. Without that information and to the extent that results from different tools, methods, and laboratories are independent, the random effects model yields an estimate of total uncertainty. Even so, any error in common among individual results is unobservable and a high priority for investigation.

Further progress is possible by the study of suspected influence quantities. Upon measurement of $q_i$, the result and its uncertainty become the $\hat{q}_i$ estimate and uncertainty. Without a measurement, however, errors of the $\hat{q}_i$ estimate are unknown, systematic, and prone to cause a discrepancy. Changing $q_i$ changes the signal, $S$, but not the measurand and so should not change $f(S, \hat{q}_i)$. This correction of the effect of $q_i$ is the purpose of $f$. Therefore, deliberate variation of influence quantities is a test of the model in $f$. As influence quantities come to light, dark uncertainty has fewer places to hide, and discrepancies among tools, methods, and laboratories decrease.

VI. CONCLUSION

Two important ideas are on a collision course, with this point of view serving as the lookout.

The prevailing idea is the promise of reducing uncertainty by combining results from different measurement processes in semiconductor manufacturing. An IEEE technology roadmap has targeted an uncertainty of ± 0.17 nm at 95 % coverage for the widths of isolated lines by the year 2028, advising that this target is possible to hit only by hybrid metrology.

The countervailing idea is the peril of combining a few results that may be inconsistent due to dark uncertainty[5]. Despite best efforts, comparisons of results from different tools, methods, and laboratories show that excess variability is a common problem. The standard process of evaluating uncertainty is only as complete as the understanding of a measurement process.

Absent a lookout, a collision seems likely as endorsements of hybrid metrology have strengthened while warnings of its limitations have weakened[6,7,11-13], motivating the uncritical combination of inconsistent results. Such analysis can cause titanic overconfidence, leading to catastrophic decisions and sinking ventures. Our study shines a light on dark uncertainty to avoid this lurking issue, and to guide combinations of results toward better—but often higher—estimates of total uncertainty.

Considering the motivation to obtain estimates of uncertainty that are both reliable and optimal, future studies could develop random effects models that benefit from the complementarity of different principles of measurement and also recognize dark uncertainty among the results. As well, hierarchical models with additional levels could account for fabrication variability within the averaging areas of different measurement processes.

Linewidth provides suitable specificity in our study, but our warning applies to other applications of hybrid metrology. Dimensional measurands such as overlay[61] can be inconsistent due to target–feature deviation, while optical, material, and chemical measurands present new challenges. Beyond semiconductor manufacturing, hybrid metrology is finding use in adjacent topics of nanostructure measurement[14-16]. Finally, the combination of measurement results is a critical issue in other manufacturing processes[62] and data fusion techniques[63].

ACKNOWLEDGMENT

Sean Kelley creatively illustrated all figures. Alan Brodie, Craig Copeland, John Gerling, John Kramar, Chris Mack, Andrew Madison, David Newton, Heather Patrick, Terri Utlaut, and Mehdi Vaez-Iravani provided helpful reviews.